\documentclass[twocolumn,twocolappendix]{aastex631}
\usepackage{color} % use \added{text}, \deleted{text}, \replaced{old}{new} instead.
\usepackage{amsmath}
\usepackage{bm}
\usepackage{mathtools}
\newcommand{\tred}[1]{#1}

\newcommand{\figdir}{./}
\newcommand{\lrsp}{Living~Rev.~Sol.~Phys.}
\newcommand{\sciadv}{Science Adv.}

\begin{document}

%%%%
%%%% Title
%%%%
\title{
  A Comprehensive Simulation of Solar Wind Formation from the Solar Interior: \\ Significant Cross-field Energy Transport by Interchange Reconnection near the Sun
}

%%%%
%%%% Authors
%%%%
\correspondingauthor{Haruhisa Iijima}
\email{h.iijima@isee.nagoya-u.ac.jp}

\author[0000-0002-1007-181X]{Haruhisa Iijima}
\affiliation{
  Institute for Space-Earth Environmental Research, Nagoya University, Furocho, Chikusa-ku, Nagoya, Aichi 464-8601, Japan
}
\affiliation{
  Institute for Advanced Research, Nagoya University, Furocho, Chikusa-ku, Nagoya, Aichi 464-8601, Japan
}

\author[0000-0002-1043-9944]{Takuma Matsumoto}
\affiliation{
  Institute for Space-Earth Environmental Research, Nagoya University, Furocho, Chikusa-ku, Nagoya, Aichi 464-8601, Japan
}
\affiliation{
  National Astronomical Observatory of Japan, 2-21-1 Osawa, Mitaka, Tokyo 181-8588, Japan
}

\author[0000-0002-6312-7944]{Hideyuki Hotta}
\affiliation{
  Institute for Space-Earth Environmental Research, Nagoya University, Furocho, Chikusa-ku, Nagoya, Aichi 464-8601, Japan
}

\author[0000-0001-7891-3916]{Shinsuke Imada}
\affiliation{
  Department of Earth and Planetary Science, The University of Tokyo, 7-3-1 Hongo, Bunkyo-ku, Tokyo, 113-0033, Japan.
}

%%%%
%%%% Abstract
%%%%
\begin{abstract}
  The physical connection between thermal convection in the solar interior and the solar wind remains unclear due to their significant scale separation. 
  Using an extended version of the three-dimensional radiative magnetohydrodynamic code RAMENS, we perform the first comprehensive simulation of the solar wind formation, starting from the wave excitation and the small-scale dynamo below the photosphere.
  The simulation satisfies various observational constraints as a slow solar wind emanating from the coronal hole boundary. 
  The magnetic energy is persistently released in the simulated corona, showing a hot upward flow at the interface between open and closed fields.  
  To evaluate the energetic contributions from Alfv\'en wave and interchange reconnection, we develop a new method to quantify the cross-field energy transport in the simulated atmosphere.
  The measured energy transport from closed coronal loops to open field accounts for approximately half of the total.
  These findings suggest a significant role of the supergranular-scale interchange reconnection in solar wind formation.
\end{abstract}

%%%%
%%%% Keywords
%%%%

%%%%
%%%% Introduction
%%%%
\section{Introduction}
\label{sec:intro}

The physical mechanisms driving the solar wind, a supersonic stream of charged particles originating from the Sun's outer atmosphere, have long been investigated \citep{2017SSRv..212.1345C}.
Two primary scenarios have been the focus of extensive research: Alfv\'en waves and interchange magnetic reconnection.
Alfv\'en waves have been investigated as an important scenario of solar wind acceleration
\citep{1977ApJ...215..942J,1986JGR....91.4111H,2005ApJ...632L..49S,2007ApJS..171..520C,2013ApJ...776..124P,2019JPlPh..85d9009C,2019ApJ...880L...2S,2021MNRAS.500.4779M}.
The interaction between the magnetic field and thermal convection excites Alfv\'en waves, propagating along the magnetic field lines into the upper atmosphere.
The dissipation of Alfv\'en waves not only heats the plasma but also creates a wave pressure gradient that accelerates the solar wind \citep{1958ApJ...128..664P,1980JGR....85.4681L}.
Conversely, the complex magnetic configuration of the solar magnetic field may produce frequent interchange magnetic reconnection between closed and open magnetic fields \citep{1992sws..coll....1A,1999JGR...10419765F,2003JGRA..108.1157F,2011ApJ...731..112A,2010ApJ...720..824C,2018ApJ...862....6C,2020ApJ...904..199W,2020ApJ...903....1Z}.
The interchange reconnection involves the fast conversion of magnetic energy into heat and waves, allowing the transfer of mass and energy across magnetic field lines.

While the Alfv\'en wave scenario satisfies various observational constraints \citep{2012SSRv..172..145C}, the interchange reconnection scenario remains an appealing explanation for the ionization degree and elemental abundance in the solar wind \citep{2016SSRv..201...55A}. The presence of magnetic switchbacks in the near-Sun solar wind \citep{2019Natur.576..237B,2019Natur.576..228K} is also attributed to the explosive energy release through the magnetic reconnection in the low corona \citep{2020ApJ...903....1Z,2020ApJ...896L..18S,2021ApJ...911...75M}. However, the in-situ formation scenario by the turbulent Alfv\'en waves can also explain some of the observational constraints of magnetic switchbacks \citep{2020ApJ...891L...2S,2021ApJ...915...52S}. Distinguishing between the Alfv\'en wave and interchange reconnection scenarios has proven difficult in both observations and theories.

Investigating these competing scenarios has been challenging due to the complexity and vast parameter space inherent in the near-surface physics of the Sun.
Turbulent motion in near-surface thermal convection is widely accepted as the direct energy source of the solar atmosphere \citep{1947MNRAS.107..211A,1989SoPh..121..271P}.
The near-surface thermal convection generates the mixed-polarity magnetic field \citep{2004Natur.430..326T,2011ApJ...737...52L} through the small-scale dynamo \citep{1993A&A...274..543P,1999ApJ...515L..39C,2007A&A...465L..43V,2014ApJ...789..132R}. The interaction of the dynamo-generated magnetic field and turbulent motion excites magnetohydrodynamic waves, which transport magnetohydrodynamic energy into the solar atmosphere \citep{1998ApJ...495..468S,2011ApJ...743..142H,2013ApJ...776L...4S,2017ApJ...848...38I,2017ApJ...834...10R}.
Empirically modeling these physical processes is difficult, often requiring numerous free parameters and arbitrary assumptions.
Additionally, the substantial scale separation between the solar surface and wind has hindered the development of self-consistent models that can describe the physical connection between them.

In this Letter, we carry out a three-dimensional radiative magnetohydrodynamic simulation of the solar wind formation starting from the solar upper convection zone.
To emulate the small-scale dynamo and wave excitations near the photosphere, we extend the established convection model \citep{1984ssdp.conf..181N,1998ApJ...499..914S,2009LRSP....6....2N}, which is almost parameter-free except the domain size or numerical resolution.
This model has been extended to study {the dynamics in the chromosphere and corona} \citep{2011A&A...531A.154G,2017ApJ...848...38I,2017ApJ...834...10R}.
Our work represents the first extension of this near-surface convection model to solar wind formation.
In this Letter, we especially focus on the energy transport process from the bottom of the corona up to the solar wind.
To achieve this, we suggest a new method for quantitatively estimating the energetic contribution of interchange reconnection in the complex simulated atmosphere (Appendix \ref{appendix:energy}).
This approach provides new insights into the competing mechanisms driving solar wind formation and offers a more comprehensive understanding of the intricate processes occurring in the Sun's outer atmosphere.

%%%%
%%%% Method
%%%%
\section{Simulation setup}
\label{sec:method}

We used an extended version of the RAMENS \citep{2016PhDT.........5I,2015ApJ...812L..30I,2017ApJ...848...38I} to simulate the solar wind acceleration from the upper convection zone.
The RAMENS solve the fully compressible magnetohydrodynamic equations, including the effects of gravity, optically thick and thin radiation, latent heat of partial ionization, and Spitzer-Harm thermal conduction in orthogonal curvilinear coordinates.
The magnetohydrodynamic equations were discretized with the energy-consistent finite difference method \citep{2021JCoPh.43510232I} to consistently account for energy exchange between the plasma and magnetic energies.
We chose the sixth-order central difference method in space and the four-step Runge-Kutta method in time.
The slope-limited diffusion scheme \citep{2014ApJ...789..132R} was adopted to stabilize the shocks and discontinuities.
No explicit viscosity or resistivity was assumed.
To resolve the thin transition region, we used the numerical broadening method \citep{2021ApJ...917...65I} to avoid underestimating the coronal density.
The optically thick radiative transfer was solved in the one-dimensional plane-parallel approximation in each vertical column to reduce the numerical cost.
The validity of the one-dimensional approximation has already been validated through comparisons with three-dimensional radiative transfer \citep{2019SciA....5.2307H}.
The semi-relativistic correction \citep{2002JCoPh.177..176G,2017ApJ...834...10R} was also used to relax the severe constraints on the time steps where the height (from the solar surface) lower than $0.1R_{\odot}$.
More details are in our previous publications \citep{2016PhDT.........5I}.

We used the localized spherical geometry accounting the global expansion of the magnetic field lines \citep{2021MNRAS.500.4779M}.
\tred{
  We define the $z$-axis to be in the radially outward direction from the Sun.
  Thus, the radial coordinate can be written as $r=z+R_\odot$.
}
The numerical domain expanded from $15$ Mm below ($z=-15$ Mm) to $29R_{\odot}$ distant from the solar surface ($z=0$).
A vertically non-uniform grid spacing was used \tred{in $z$-direction} to reduce the numerical cost.
The vertical grid size was set to 228 km at the bottom boundary, $48.2$ km at the surface, and gradually increased to $9.65$ Mm at the top boundary.
The super-radial expansion factor was set similarly to \cite{2021MNRAS.500.4779M} but uses the value $4$ at the top boundary.
We forced the purely radial expansion (without super-radial expansion) below the surface.
The horizontal size of the numerical domain was \(48 \times 48\) Mm$^2$ at the surface, sufficiently including several supergranular-scale $\approx 30$ Mm convection cells \citep{1962ApJ...135..474L,1964ApJ...140.1120S}.
The corresponding horizontal domain size at the top boundary is $4.1R_\odot$.
The horizontal grid size was constant for each height and set to $188$ km at $z = 0$ and $11.3$ Mm at the top boundary.
The entire numerical domain is covered by $(N_x,N_y,N_z)=(256,256,3072)$ grids.

The bottom boundary was open for plasma flow, allowing thermal and magnetic energy transports into the domain. The total (gas plus magnetic) pressure was kept constant at the bottom boundary, allowing weak perturbations to suppress the standing sound waves in the convection zone. The top boundary was open for plasma flow and the outward Els\"asser variables. The inward Els\"asser variables were set to zero. The details of the top boundary condition do not strongly affect the solution below as the solar wind travels faster than Alfv\'en and sound waves at this height. Periodic boundary conditions were assumed in each horizontal direction.

The simulation started from a snapshot of the sufficiently relaxed magneto-convection simulation with a horizontally averaged signed vertical magnetic flux (flux imbalance) of 6 G at the photospheric level. The small-scale dynamo in the numerical domain amplified the magnetic energy in the convection zone and generated mixed-polarity magnetic patterns in the photosphere. Once the volume-integrated magnetic energy in the convection zone had saturated, we imposed a solar-like one-dimensional stratification and potential (current-free) magnetic field. Here we used the one-dimensional empirical atmosphere \citep{2008ApJS..175..229A} with a smooth transition to the isothermal Parker solution \citep{1958ApJ...128..664P} at $T= 1$ MK. The simulation was integrated for 72 h. After the initial transients, the simulation appeared to relax, considering the travel times of Alfv\'en and sound waves. We mainly analyzed the last 48 h in this paper.

%%%%
%%%% Results
%%%%
\section{Results}
\label{sec:results}

%%%%
\subsection{Overall structure}
\label{subsec:overall}
%%%%

\begin{figure*}[t!]
  % \epsscale{0.7}
  \plotone{{\figdir}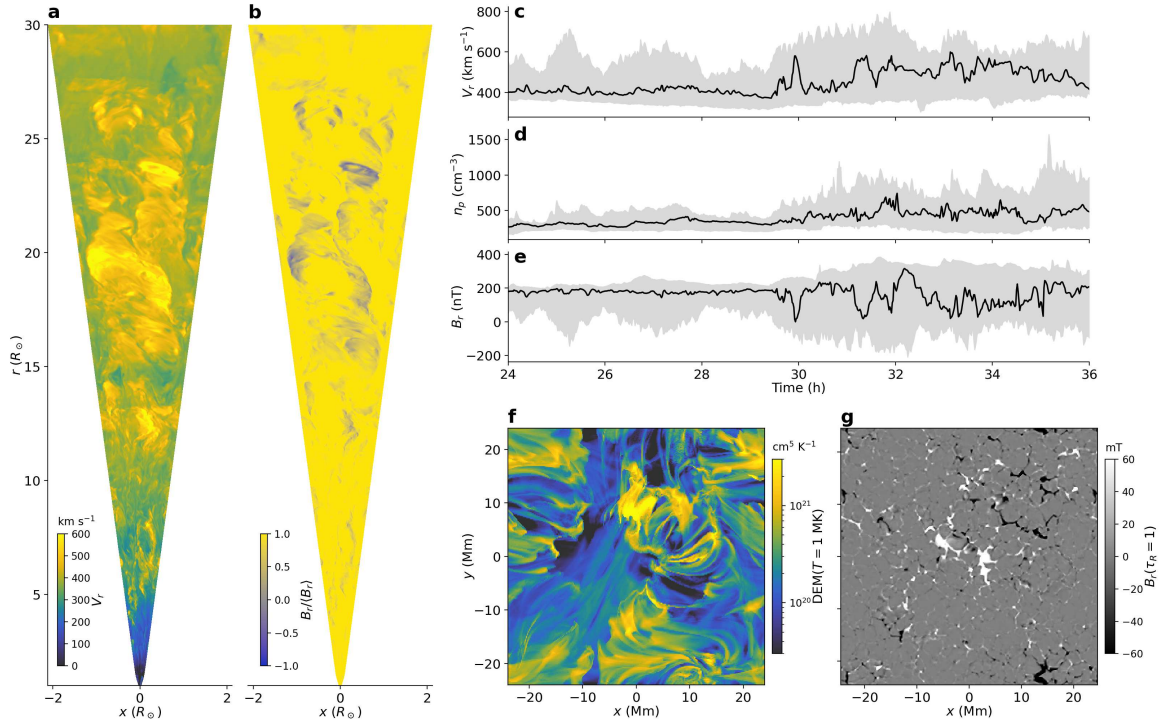}
  \caption{
    Overview of the simulated solar atmosphere.
    (a--b) Radial velocity field and radial magnetic field (normalized by its horizontal average), respectively, on a $xr$-plane at $y = 0$, \tred{where $r=z+R_\odot$.}
    (c--e) Radial velocity field, proton number density, and radial magnetic field measured on a $xy$-plane at $r = 29R_\odot$, respectively. In each panel, the point value at $(x, y) = (0, 0)$ is shown as a solid black line, and the gray-shaded region represents the range of the maximum and minimum values on the same $xy$-plane.
    (f) Differential emission measure at $T = 1$ MK observed from the top of the simulation box.
    (g) Radial magnetic field in the solar photosphere (at the optical depth unity).
    In panels (a,b,f,g), all variables are measured at time = 29 h.
  }
  \label{fig:plwhole}
\end{figure*}

Figure \ref{fig:plwhole} shows an over view of the simulated solar atmosphere.
The supersonic solar wind and million-degree corona are generated spontaneously by the magneto-convection within the simulation domain.
In a steady state, the mass-loss rate of the solar wind is roughly proportional to the magnitude of the energy input \citep{1994ApJ...436..390S,1995JGR...10021577H,2011ApJ...741...54C}, thus can be used as a measure of the energy input to the solar wind. The simulated mass-loss rate was determined as $3\text{-}10 \times 10^{-14}M_{\odot}$/yr, which is similar to but slightly larger than the typical observed value \citep[$2\text{-}3 \times 10^{- 14} M_{\odot}$/yr;][]{2017SSRv..212.1345C}. This discrepancy may relate to the strong photospheric magnetic energy (see the discussion in Sec. \ref{sec:discussion}). As the observed photospheric magnetic field is highly non-uniform in space and time, our simulation should realize at least a part of the physical processes and parameter space of the real Sun.

The solar wind is highly structured in space and time, sometimes showing polarity reversal of the radial magnetic field (Fig. \ref{fig:plwhole}a--e), known as magnetic switchbacks \citep{2019Natur.576..237B,2019Natur.576..228K}.
According to an observational study by Parker Solar Probe \citep{2019Natur.576..237B}, the fraction of magnetic switchbacks (filling factor of $B_r<0$) comprise approximately 6\% of the temporal duration.
Conversely, the simulated filling factor was 0.33\% in the time range of 24--72 h, which is an order of magnitude smaller than the observed value. The in-situ enhancement of the magnetic switchbacks \citep{2020ApJ...891L...2S,2021ApJ...915...52S}, which strongly depends on the numerical resolution, may resolve this discrepancy.

This dynamic solar wind flows from the underlying corona, which shows
\tred{
  a complex thermodynamic structure with mixed dark open field regions and bright coronal loops (Fig. \ref{fig:plwhole}f; see also Fig. \ref{fig:plmr}).
}
% This dynamic solar wind flows from the underlying corona, which has a complex magnetic topology with mixed open and closed magnetic field lines (Fig. \ref{fig:plwhole}f).
The emerging radiation from the corona has been used as a measure of the coronal heating rate \citep{1977ARA&A..15..363W}. We calculated the differential emission measure \citep{2018LRSP...15....5D} near ${T} = 1$ MK and compared it with the observations \citep{1978ApJS...37..485V} stored in the CHIANTI atomic database v.9 \citep{2019ApJS..241...22D}. The simulated value was $3.9 \times 10^{20}$ cm$^5$ K$^{-1}$, comparable to the typical parameter in the quiet region ($3.1 \times 10^{20}$ cm$^5$ K$^{-1}$) and larger than that in the coronal hole ($3.9 \times 10^{19}$ cm$^5$ K$^{-1}$). In this sense, our model realizes a slow solar wind blowing from coronal hole boundary near a quiet region rather than a fast solar wind from a dark polar coronal hole.

The simulated atmosphere is powered by the turbulent magneto-convection. The small-scale dynamo in the convection zone amplifies the magnetic energy and creates mixed-polarity magnetic patterns in the photosphere \citep[Fig. \ref{fig:plwhole}g;][]{1993A&A...274..543P,1999ApJ...515L..39C,2007A&A...465L..43V}.
The strength of the small-scale magnetic field is a vital free parameter that affects the energy input to the solar atmosphere. The magnetic energy at the solar surface is determined by the balance between the small-scale dynamo, downward transport of magnetic energy \citep{1993A&A...274..543P,2003ASPC..286..121S}, and inflow of magnetic energy through the bottom boundary from the deeper solar interior \citep{2014ApJ...789..132R}. In our model, the simulated unsigned vertical magnetic flux at the optical depth unity was 51.2 G (5.12 mT), slightly smaller than the typical internetwork magnetic field strength of 85 G \citep[8.5 mT;][]{2016A&A...594A.103D,2018ApJ...863..164D}. Considering the different spatial resolutions of our model and the previous studies \citep{2014ApJ...789..132R,2016A&A...594A.103D,2018ApJ...863..164D}, we speculate that the realized photospheric magnetic energy is within a reasonable range in the quiet Sun or in the coronal hole.

%%%%
\subsection{Time variation}
\label{subsec:variation}
%%%%

\begin{figure}[t!]
  % \epsscale{0.7}
  \plotone{{\figdir}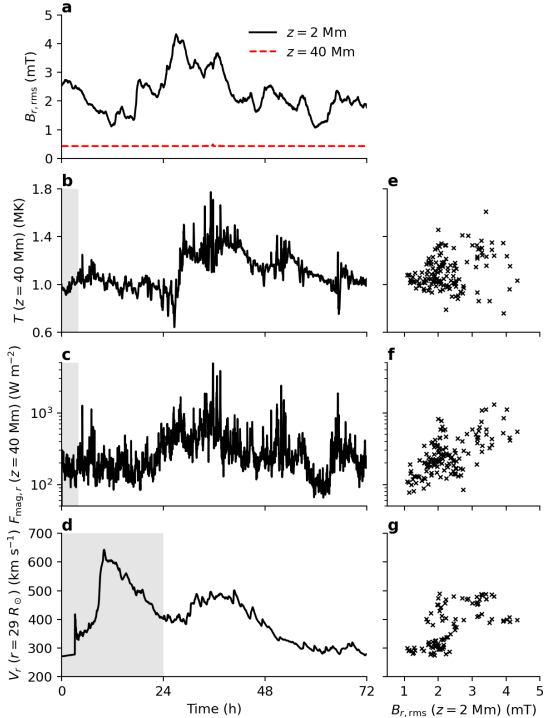}
  \caption{
    Time variations in the solar atmosphere with magnetic activity near the surface.
    (a) Root-mean-square of the radial magnetic field in a horizontal plane at $z=2$ Mm (base of the corona, solid black line) and $z=40$ Mm (middle corona, dashed red line).
    (b,c) Horizontal average of the plasma temperature and radial magnetic
energy flux, respectively, measured at $z=40$ Mm.
    (d) Horizontal average of the radial velocity field at $r =
29R_{\odot}$.
    (e--g) Dependencies of the coronal temperature, coronal magnetic energy flux, and solar wind speed, respectively, on the root-mean-square of the magnetic field at $z=2$ Mm.
    % Time delays of 1, 1, and 6 h were accounted for to mimic the physical delays in the response due to wave propagation and thermal relaxation.
    The shaded region in panels (b)--(d) is possibly affected by the initial condition and is thus excluded from the scatter plots (e)--(g) and other analyses.
  }
  \label{fig:pl0d_time}
\end{figure}

The simulated atmosphere varies on daily time scales (Fig. \ref{fig:pl0d_time}). Randomness of the thermal convection and small-scale dynamo produces the variation in the realized magnetic field strength at the base of the solar corona (Fig. \ref{fig:pl0d_time}a, solid black line). In contrast, the magnetic field strength at \emph{z} = 40 Mm is almost constant in time, as most magnetic flux is open (connected to the solar wind). The coronal temperature (Fig. \ref{fig:pl0d_time}b), vertical component of magnetic energy flux (Fig. \ref{fig:pl0d_time}c), and speed of the solar wind (Fig. \ref{fig:pl0d_time}d) show time variations similar to the near-surface magnetic energy (Fig. \ref{fig:pl0d_time}e-g).
% To mimic the travel speed of information through Alfv\'en and sound waves, we impose a time delay between the measured heights of each variable. After incorporating the time delay, the simulated atmosphere depends on the magnetic energy at the bottom of the solar corona (Fig. \ref{fig:pl0d_time}e-g).

The positive dependence of the coronal temperature (and the energy dissipation in the corona) on the surface magnetic activity may be explained by (1) the random shuffling of the magnetic field lines \citep{1989SoPh..121..271P,1998Natur.394..152S}, (2) the Alfv\'en waves traveling the expanding magnetic flux tube \citep{1984SoPh...91..269H,2012A&A...538A..70V}, or (3) the size distribution of closed loops \citep{1978ApJ...220..643R,2004SoPh..225..227W}. To satisfy energy balance in the corona, the higher coronal temperature increases the mass density and inertia of the flowing plasma, interfering with the acceleration of the solar wind \citep{1980JGR....85.4681L}. Thus, the time variation of the solar wind speed should be explained by variations in the magnetic energy flux transported to the solar wind (Fig. \ref{fig:pl0d_time}c, f).

The time variation of the magnetic energy flux at $z = 40$ Mm is not easily explained by the field-aligned energy transport by Alfv\'en waves \citep{1984SoPh...91..269H,2012A&A...538A..70V}. Assuming that the velocity amplitude of excited Alfv\'en waves in the photosphere and the amount of wave dissipation during the propagation are independent of the surface magnetic energy, the magnitude of the magnetic energy flux should be proportional to the local magnetic field strength. However, as most of the magnetic flux is open and uniformly distributed above $z = 40$ Mm (Fig. \ref{fig:pl0d_time}a, dashed red line), the Alfv\'en wave model predicts a temporally constant magnetic energy flux at this height, which contradicts the measurements shown in Fig. \ref{fig:pl0d_time}f. We expect another process that significantly contributes to energy transport.

%%%%
\subsection{Persistent energy release in the low corona}
\label{subsec:energy_release}
%%%%

\begin{figure*}[t!]
  % \epsscale{0.7}
  \plotone{{\figdir}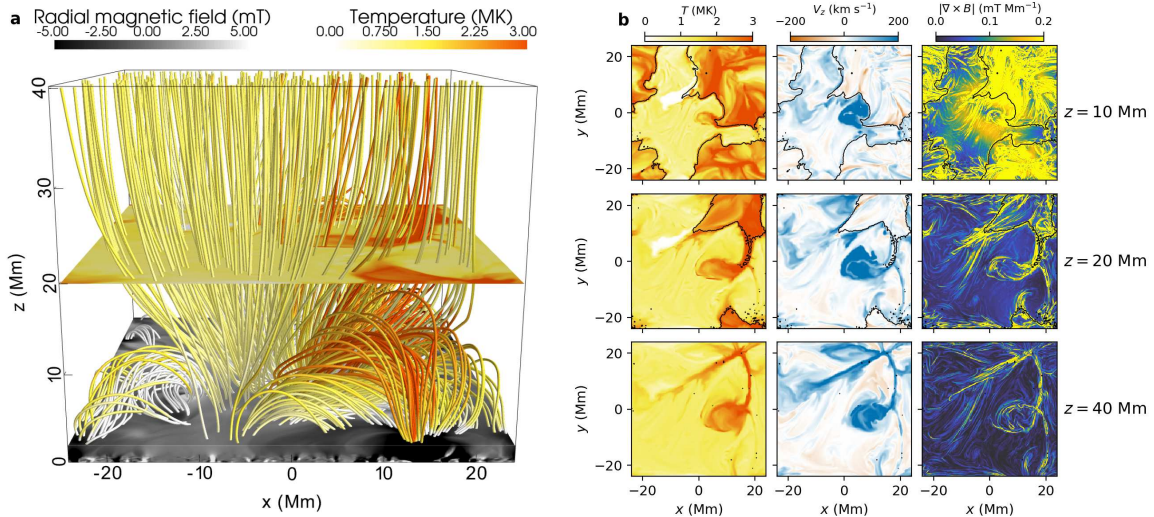}
  \caption{
    Magnetic and thermodynamic structures in the low corona.
    (a) Three-dimensional structure of the magnetic field lines. The line color indicates the plasma temperature. Horizontal slices of the radial magnetic field at \emph{z} = 2 Mm and plasma temperature at \emph{z} = 20 Mm are also shown.
    (b) Horizontal slices of
    (left) the plasma temperature $T$,
    (middle) the vertical velocity $V_z$, and 
    (right) the magnitude of current density $\left|\nabla\times\bm{B}\right|$
    at different heights.
    The thin solid lines indicate the boundary between the open and closed magnetic regions.
    Most of the field lines are open at $z = 40$ Mm.
  }
  \label{fig:plmr}
\end{figure*}

Looking closely at the low corona, we find that magnetic energy is persistently released at the boundary of the open and closed magnetic regions.
The small-scale dynamo in the convection zone generates a complex coronal magnetic structure.
Fig. \ref{fig:plmr}a shows the open and closed magnetic field lines that are mixed and closely spaced.
The field lines with high temperatures can be observed not only in the closed field region but in the open field region near the closed loops.
Near the open field region with hot plasma, a fast upward flow ($\ge 200$ km s$^{-1}$) appears around $(x,y) = (5, 0)$ Mm.
This upward mass flow converges into hot and thin boundary layers at $z = 40$ Mm and supplies mass and energy to the solar wind above.
The electric current exhibits a spatially complex structure, possibly due to the shearing motion by the photospheric convection.
The strong electric current in the closed field region sometimes penetrates the open field region, forming the observed hot and fast upflow.
% Those results, hot and fast upflow with complex current structure near the open-closed  imply the magnetic energy release through magnetic reconnection.
% 
Similar magnetic energy release continues during the simulation run.
This hot and fast plasma upflow with the strong \tred{electric} current near the boundary of open and closed field regions should be a realization of the interchange reconnection scenario of the solar wind \citep{1992sws..coll....1A,1999JGR...10419765F,2003JGRA..108.1157F}.

%%%%
\subsection{Cross-field energy transport}
\label{subsec:energy_transport}
%%%%

\begin{figure}[t!]
  % \epsscale{0.5}
  \plotone{{\figdir}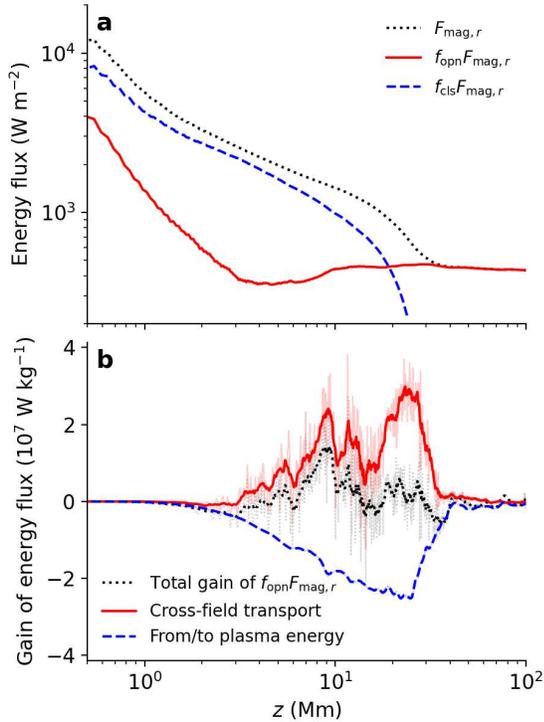}
  \caption{
    Strong energy transport from the closed coronal loops into the solar wind.
    (a) Vertical profile of the horizontally averaged magnetic energy flux (black dotted line). Energy fluxes in the open (solid red line) and closed (dashed blue line) field regions are also shown.
    (b) Gain and loss of the magnetic energy flux per unit mass (roughly corresponding to the cooling and heating rates of plasma, respectively) in the open-field region. Thick and thin light lines show the vertically low-pass filtered and unfiltered data, respectively.
  }
  \label{fig:pl1d_fmag}
\end{figure}

Neither observational nor theoretical studies have quantitatively
revealed the relative importance between the Alfv\'en wave and
interchange reconnection scenarios on the solar wind formation \citep{2017SSRv..212.1345C}. The Alfv\'en wave scenario \citep{1977ApJ...215..942J,1986JGR....91.4111H,2005ApJ...632L..49S,2007ApJS..171..520C,2019ApJ...880L...2S} satisfies various observational constraints \citep{2012SSRv..172..145C}, whereas the interchange reconnection scenario \citep{1992sws..coll....1A,1999JGR...10419765F,2003JGRA..108.1157F,2010ApJ...720..824C,2018ApJ...862....6C,2011ApJ...731..112A,2020ApJ...903....1Z} has maintained persistent interest as a natural explanation for the variability, the ionization degree, and the elemental abundance in the solar wind \citep{2016SSRv..201...55A}. Because our model includes contributions from both scenarios, we can evaluate the amount of energy input by the Alfv\'en wave and interchange reconnection scenarios.

We use the amount of cross-field energy transport (Poynting flux) to measure energy input by the interchange reconnection scenario (Appendix \ref{appendix:energy}).
In most models, the Alfv\'en waves basically travels along the magnetic field lines in the magnetized atmosphere, carrying the magnetic energy into solar wind.
Thus, if we could estimate the energy transport across the interface between the open and closed fields, it is also possible to evaluate the relative contribution of Alfv\'en wave and interchange reconnection in the simulated atmosphere.  

Fig. \ref{fig:pl1d_fmag}a shows the vertical profile of the magnetic energy flux along open and closed field regions. Most of the magnetic energy is transported into the closed field region below $z=10$ Mm rather than into the open field region connected to the solar wind, reflecting the area fraction of each region in the low corona. As the number of closed field lines decreases with height, the energy flux in the closed region becomes zero at $z = 30$ Mm. In contrast, the energy flux in the open field region is slightly enhanced from \emph{z} = 3 Mm to $z = 30$ Mm.

We then separated the contributions to the vertical variation of magnetic energy flux in the open field region. As shown in Fig. \ref{fig:pl1d_fmag}b, a significant amount of magnetic energy is transferred from the closed to open field regions (denoted as cross-field transport, see Eq. \ref{eq:cf_energy}). Most of the magnetic energy transferred from the closed magnetic region is dissipated and heats the coronal plasma in the open field region. The net energy gain per magnetic flux density (a measure of energy gain invariant from the expansion of magnetic flux tube; see Eq. \ref{eq:cf_energy_intz}) from the closed magnetic region between $z = 2$ Mm and $z = 40$ Mm is $9.38 \times 10^{5}$ W m$^{-2}$ T$^{-1}$, comparable to the energy transported directly from the lower atmosphere ($9.93 \times 10^{5}$ W m$^{-2}$ T$^{-1}$). In other words, approximately half of the total solar wind energy comes from the closed coronal loops.

%%%%
%%%% Discussion
%%%%
\section{Discussion}
\label{sec:discussion}

\begin{figure}[t!]
  % \epsscale{0.6}
  \plotone{{\figdir}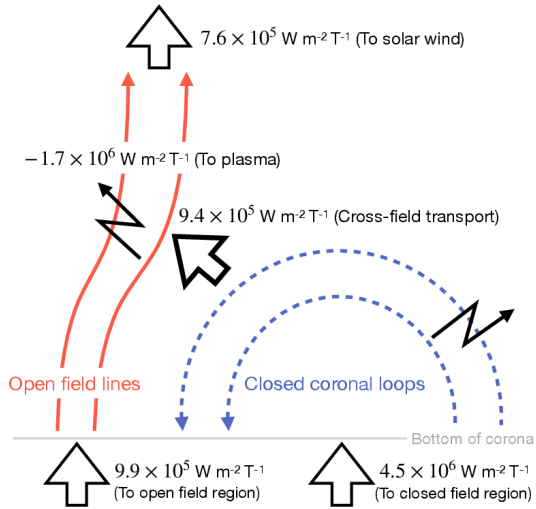}
  \caption{
    Schematic illustration of energy transport in the solar corona.
    Field-aligned transport supplies half of the total energy input in the open field region. The remaining half comes from the closed field region through the interactions between open and closed magnetic field lines.
  }
  \label{fig:cartoon}
\end{figure}

The evaluated energy transport process along and across the magnetic field lines in the simulated corona is summarized in Fig. \ref{fig:cartoon} (see also Appendix \ref{appendix:energy} for details).
Approximately half of the energy input to the solar wind is transferred from closed coronal loops to the open field region.
The existence of hot plasma, confined upward flow, and strong electric current at the interface between open and closed fields (Fig. \ref{fig:plmr}) implies that the cross-field energy transport is caused by the interchange reconnection in low corona.
Thus, our simulation can be interpreted as a realization of the interchange reconnection scenario of solar wind on the supergranular-scale \citep{1992sws..coll....1A,1999JGR...10419765F,2003JGRA..108.1157F,2010ApJ...720..824C,2018ApJ...862....6C}.

We noticed that the emergence of magnetic flux is not as evident in our simulated corona, in contrast to the previous models of interchange reconnection scenario \citep{2003JGRA..108.1157F,2010ApJ...720..824C,2020ApJ...904..199W}. Instead, the horizontal braiding motion in the low corona appeared to cause the interchange reconnection at the interface between open and closed fields. This difference may resolve smaller energy input than required for solar wind acceleration, as reported by some interchange reconnection models \citep[e.g.,][]{2010ApJ...720..824C}. These speculation may help update the picture of interchange reconnection scenario of the solar wind and warrant further investigation.

Our results do not exclude the potential contribution of interchange reconnection in larger-scale (quasi-)separatrix layers, like helmet streamers or pseudostreamer boundaries \citep[e.g.,][]{2011ApJ...731..112A}. As we assume a supergranular-scale horizontal domain, the model cannot mimic the global-scale interchange reconnection.

The dependence on the numerical grid size will be observed, especially near the photosphere (where a single granulation is resolved only 5--10 grids).
The rate of magnetic reconnection in the numerical simulation also affected by the numerical resistivity in the model, until the plasmoid reconnection becomes dominant.
Therefore, the insufficient numerical resolution may affect the amount of cross-field energy transport.
We are going to investigate such numerical artifacts in future studies.

Although we focus only on energy transport in the corona, high plasma beta, complex magnetic structure, and high nonlinearity (e.g., shock waves) in the chromosphere may easily lead larger amount of cross-field transport of mass and energy. As the numerical resolution may not be sufficient in our simulated chromosphere, we did not show a detailed analysis of energy transport in the chromosphere. This point should be investigated in high-resolution simulations.

The energetic contribution from the closed region showed a dependence on the photospheric magnetic energy (that relates to the number of coronal loops).
Given the spatial resolution of observed synoptic magnetic maps and the prediction models focusing on large-scale magnetic structures, the small-scale magnetic structure may be a missing factor in the large dispersion observed in empirical solar wind models \citep{2015SpWea..13..154R}.

Although the simulated atmosphere successfully reproduced various observational constraints (Sec. \ref{subsec:overall}) as a slow solar wind blowing from a coronal hole boundary, the mass-loss rate was larger than the average solar wind value. The unsigned magnetic flux in the photospheric layer was also slightly over-enhanced.
These discrepancies may be caused by insufficient numerical resolution and the magnetic bottom boundary condition.
The realized photospheric magnetic energy by the small-scale dynamo simulations is affected by the strength of the magnetic flux recirculation \citep{2014ApJ...789..132R,2018ApJ...859..161R}. Similar dependence on the deep recirculation (modeled as the magnetic bottom boundary condition) was also observed in our simulation.
In this context, our results imply that the small-scale dynamo and the turbulent magnetic energy in the deep convection zone may control the nature of the solar wind.

% %%%%
% %%%% Conclusion
% %%%%
% \section{Conclusion}
% \label{sec:conclusion}

% We have presented the first comprehensive model of solar wind formation starting from magneto-convection in the solar interior. Within this model, the cross-field energy transport caused by the interchange reconnection between the closed and open magnetic field lines in low corona contribute a significant portion of the solar wind energy. This result infers a potential impact of the small-scale dynamo in the solar interior on the nature of solar wind.

%%%%
%%%% Acknowledgements
%%%%
%\begin{acknowledgments}
\section*{}
  The authors would like to thank the anonymous referee for the constructive comments, which improved the quality of the paper.
  % funding
  H.I. was supported by JSPS KAKENHI grant no. JP19K14756, JP21H01124, and JP23K13144.
  H.I. was also supported as the young researcher units for the advancement of new and undeveloped fields of the Institute for Advanced Research (Nagoya University) under the Program for Promoting the Enhancement of Research Universities.
  T.M. was supported by JSPS KAKENHI grant no. JP23K03456.
  H.H. was supported by JSPS KAKENHI grant no. JP20K14510, JP21H04492, JP21H04497, and JP23H01210.
  S.I. was supported by JSPS KAKENHI grant no. JP20K20943.
  % Fugaku
  This work used computational resources of supercomputer Fugaku provided by the RIKEN Center for Computational Science through the HPCI System Research Project (Project ID: hp210144).
  This work was supported by MEXT as a Program for Promoting Researches on the Supercomputer Fugaku (grant no. J20HJ00016, JPMXP1020230406, and JPMXP1020230504).
  % CfCA
  A part of numerical computations was performed on Cray XC50 at Center for Computational Astrophysics, National Astronomical Observatory of Japan.
  % Flow
  A part of numerical computations were also performed on the Flow supercomputer system at the Information Technology Center, Nagoya University, through the "Computational Joint Research Program (Collaborative Research Project on Computer Science with High-Performance Computing)" at the Institute for Space-Earth Environmental Research, Nagoya University.
  % CIDAS
  Analysis of the numerical simulations were performed in the Center for Integrated Data Science, Institute for Space-Earth Environmental Research, Nagoya University through the joint research program.
  % Discussions
  We are grateful for productive discussions with K. Kusano, T. Suzuki, M. Shoda, K. Iwai, D. Shiota, and many collaborators.
%\end{acknowledgments}

%%%%
%%%% Appendix
%%%%
\appendix
% \restartappendixnumbering

%%%%
\section{Estimation of cross-field energy transport}
\label{appendix:energy}
%%%%

Energy is transported along and across magnetic field lines in the solar atmosphere. The Alfv\'en wave scenarios assume energy transport along the magnetic field lines. In contrast, magnetic reconnection induces topological changes in the magnetic field lines, allowing energy transport from closed coronal loops to open field lines connected to the solar wind. Thus, the cross-field energy transport can quantitatively determine the violation of the Alfv\'en wave scenario. The vertical variation of the magnetic energy (Poynting) flux in the open field region can be written as
\begin{align}
  \label{eq:cf_energy}
  &\frac{
    \partial
    \left<
    f_{\rm opn}F_{{\rm mag},r}
    \right>
  }{
    \partial r
  }
  =
  \left<\left(E_{\rm mag}\frac{\partial}{\partial t}+{\bf F}_{\rm mag}\cdot\nabla\right)f_{\rm opn}\right>
  \nonumber\\
  &\hspace{2em}
  -\left<f_{\rm opn}\left(W_{\rm L}+Q_{\rm res}\right)\right>
  -\left<\frac{\partial (f_{\rm opn}E_{\rm mag})}{\partial t}\right>
\end{align}
where $f_{\rm opn}$, $W_{\rm L}$, $Q_{\rm res}$, and $E_{\rm mag}$ indicate the area filling factor of the open magnetic flux (connected to the solar wind), the work done by the Lorentz force, the resistive heating rate produced by numerical diffusion, and the magnetic energy density, respectively. The bracket represents integration over the horizontal domain. The first term on the right-hand side describes the magnetic energy transport across the boundary of the open and closed magnetic regions. The second term on the right-hand side describes the energy conversion from/to the thermal and kinetic energies in the open magnetic region. The last term is the contribution from the time variation of magnetic energy.

The filling factor $f_{\rm opn}$ is calculated by tracing the magnetic field lines in each snapshot. At each grid point, the filling factor was set to unity if a magnetic field line through the grid cell reached $z = 200$ Mm, where most of the magnetic flux was connected to the solar wind in our simulation.
If open field lines did not cross a grid point, its filling factor was set to zero.

The error in the field line tracing can be estimated from the conservation of the vertical magnetic flux. As the horizontal boundary conditions are periodic, the horizontally integrated vertical magnetic flux is conserved during the time integration. The vertical magnetic flux in the open field region is then balanced to the horizontally averaged vertical magnetic flux as
$
\left<B_r\right>
=
\left<f_{\rm opn}B_r\right>
$.
% through the following identity
% \begin{equation}
%   \label{eq:flux_consv}
%   \left<B_r\right>
%   =
%   \left<f_{\rm opn}B_r\right>
%   \text{.}
% \end{equation}
In a snapshot taken at time = 72 h, the relative error in this magnetic flux conservation was less than 14\% (6\%) above $z=2$ Mm (4 Mm). The error level did not affect the conclusions of our study. To confirm the robustness of the measurement, we changed the number of magnetic field lines traced in each snapshot and the step size along one field line during the tracing.

The energy flux per magnetic flux density is invariant along a magnetic field line if the energy gains are negligible. The net contribution within the height interval between $z=z_0$ and $z=z_1$ can be obtained by integrating Eq. \ref{eq:cf_energy} along the $z$-direction
\begin{align}
  \label{eq:cf_energy_intz}
  &\left(
    \frac{
      \left<f_{\rm opn}F_{{\rm mag},r}\right>
    }{
      \left<f_{\rm opn}B_r\right>
    }
  \right)_{z=z_1}
  -
  \left(
    \frac{
      \left<f_{\rm opn}F_{{\rm mag},r}\right>
    }{
      \left<f_{\rm opn}B_r\right>
    }
  \right)_{z=z_0}
  \nonumber\\
  =&
  \frac{1}{\left<f_{\rm opn}B_r\right>}
  \int_{z_0}^{z_1}
  \left[
  % \right.\nonumber\\&\left.
    \left<
      \left(E_{\rm mag}\frac{\partial}{\partial t}+{\bf F}_{\rm mag}
      \cdot\nabla\right)f_{\rm opn}
    \right>
  \right.\nonumber\\&\left.
    -\left<
      f_{\rm opn}\left(
        W_{\rm L}+Q_{\rm res}
      \right)
    \right>
  % \right.\nonumber\\&\left.
    -\left<
      \frac{\partial (f_{\rm opn}E_{\rm mag})}{\partial t}
    \right>
  \right]
  {\rm d}z
  \text{,}
\end{align}
assuming the conservation of the horizontally integrated magnetic flux along the vertical axis
$
\left<f_{\rm opn}B_r\right>_{z=z_1}
=\left<f_{\rm opn}B_r\right>_{z=z_0}
$.

% \begin{equation}
%   \label{eq:flux_z0z1}
%   \left<f_{\rm opn}B_r\right>_{z=z_1}
%   =\left<f_{\rm opn}B_r\right>_{z=z_0}
%   \text{.}
% \end{equation}

If the energy gains (right-hand-side terms of Eq. \ref{eq:cf_energy}) can be neglected, the energy flux per magnetic flux density becomes invariant in the vertical direction. \tred{We used the sixth-order finite difference operator to evaluate the spatial derivatives and the second-order Crank-Nicolson method to evaluate the temporal derivatives.} Evaluating Eq. \ref{eq:cf_energy_intz} between $z_0=2$ Mm and $z_1=40$ Mm, the cross-field energy transport was obtained as $9.38 \times 10^{5}$ W m$^{-2}$ T$^{-1}$, approximately half of the total energy input to the open field region (Fig. \ref{fig:cartoon}). The input from the chromosphere (second term in the left-hand side of Eq. \ref{eq:cf_energy_intz}) was $9.93 \times 10^{5}$ W m$^{-2}$ T$^{-1}$, whereas the outgoing energy flux (first term in the left-hand side of Eq. \ref{eq:cf_energy_intz}) was $7.55 \times 10^{5}$ W m$^{-2}$ T$^{-1}$. The energy converted into thermal and kinetic energies was $1.69 \times 10^{6}$ W m$^{-2}$ T$^{-1}$. The magnetic energy decreases between 24 h and 72 h (Fig. \ref{fig:pl0d_time}) was $5.44 \times 10^{4}$ W m$^{-2}$ T$^{-1}$, smaller than the other terms. Finally, the residual energy gain was determined as $4.60 \times 10^{5}$ W m$^{-2}$ T$^{-1}$, mainly coming from a low time cadence of 1 h in our dataset. Note that the last term in Eq. \ref{eq:cf_energy} can be evaluated accurately after time averaging (as the magnetic energy difference between time = 24 h and 72 h) independently of cadence in the dataset. Thus, assuming that the residual comes from the error in the time derivatives of the first term in Eq. \ref{eq:cf_energy_intz}, a dataset with a higher cadence can increase the amount of cross-field transport. In other words, our evaluated cross-field transport may be a lower bound.

%%%%
%%%% Bibliography
%%%%
% \bibliographystyle{aasjournal}
% \bibliography{reference}

%%%%
%%%% List of changes
%%%%
%% uncomment this line to list changes by \added{}, \deleted{}, or \replaced{}{}.
% \listofchanges

%% end document
\end{document}